# Inhibition of whisker growth by crafting more decomposition-resistant Ti$_2$SnC MAX phase through vanadium solid solution

Haifeng Tang[1], Xiaodan Yin[1], Peigen Zhang[*,1], Victor Karpov[2], Vamsi Borra[3], Zhihua Tian[1], Jianxiang Ding[*,4], and ZhengMing Sun[*,1]

1 School of Materials Science and Engineering, Southeast University, Nanjing 211189, China

2 Department of Physics and Astronomy, University of Toledo, Toledo, Ohio 43606, USA

3 Electrical and Computer Engineering Program, Rayen College of Engineering, Youngstown State University, Youngstown, OH, 44555, USA

4 School of Materials Science and Engineering, Anhui University of Technology, Ma'anshan 243002, China

## Abstract

The exceptional synergy of ceramic and metallic properties within MAX phases positions them as highly promising for a wide array of applications. However, the stability of MAX phases has been called into question due to the phenomenon of their A-site metal whisker growth. Herein, we have significantly mitigated tin whisker growth in Ti$_2$SnC by incorporating vanadium solutes at its M-site. With an increase in vanadium concentration, there is a marked reduction in the degree of decomposition of the M-site solid solution when subjected to the same level of externally destructive treatments, thereby inhibiting whisker proliferation. Both experimental outcomes and theoretical calculations reveal that the vanadium solid solution augments the hardness, Pugh's ratio, and Poisson's ratio of Ti$_2$SnC. This solid solution exhibits enhanced mechanical strength and toughness, as evidenced by its electronic density of states and bulk modulus. These findings suggest that the incorporation of vanadium atoms introduces stronger V-C and V-Sn bonds, thus significantly bolstering the resistance of MAX phases against decomposition and effectively curtailing whisker growth. Additionally, the phenomenon reported in this paper also conforms to the electrostatic theory of whisker growth. This work for the first time achieves the suppression of A-site whisker growth through an M-site solid solution, thereby extending their potential for applications where durability and reliability are paramount.

[*] Corresponding authors. *E-mail addresses:* zhpeigen@seu.edu.cn (P. Zhang), jxding@ahut.edu.cn (J. Ding), and zmsun@seu.edu.cn (Z.M. Sun)

# 1 Introduction

MAX phases are a class of ternary layered transition metal carbides or nitrides. Their general formula is $M_{n+1}AX_n$ (where M represents early transition metal elements, A represents main group elements, X represents carbon or nitrogen, and n typically ranges from 1 to 3) [1]. The atomic structure of MAX phases can be visualized as alternating stacking of $M_6X$ octahedral structures dominated by covalent bonds and $M_6A$ trigonal prismatic structures dominated by metallic bonds [2]. This unique hybrid layered structure imparts high damage tolerance [3], high electrical conductivity [4], and resistance to thermal shocks [5].

Due to the excellent properties of MAX phases, MAX phases hold significant potential for applications in various fields such as high-temperature electromagnetic shielding, high-speed train pantographs, electric contact materials [6, 7], high-temperature load-bearing structures [8], and the nuclear industry [9]. However, in 1999, Barsoum et al. observed that polished samples of $Cr_2GaN$ exhibited visible Ga whiskers on their surfaces [10]. The phenomenon of whisker growth in MAX phases, such as Sn [11], In [12], Ga [13], and ZnO [14] whiskers, has been widely reported. This phenomenon bears a striking resemblance to the spontaneous growth of metal whiskers observed in electronic systems [15], a longstanding issue in the electronics industry [16]. Similarly, the phenomenon of whisker growth raises concerns about the structural stability and practical application of MAX phases.

Solid solution alloying, as one of the primary strategies for designing the structure and properties of MAX phases, has the capability to enhance the performance of MAX phases through the solid solution effect and the cocktail effect. Qiao et al. employed the solid-state sintering method to fabricate high-entropy $(Mo_{0.25}Cr_{0.25}Ti_{0.25}V_{0.25})_3AlC_2$, which exhibits exceptional electromagnetic microwave absorption properties. Solubilization strategies further enhance the mechanical performance of MAX phases [17]. Lapauw et al. achieved a significant enhancement in the fracture toughness of the solid solution by introducing Zr into the M-site of $Nb_4AlC_3$, thereby augmenting the material's human-enhanced characteristics [18]. Qu et al. demonstrated a notable enhancement in the hardness of $Ti_3SiC_2$ by solubilizing a certain amount of Zr into its M-site [19]. Additionally, Wei et al. synthesized $(Ti_{1-x}V_x)_2GaC$ solid solution, which showed increased hardness and wear resistance compared to the original $Ti_2GaC$ [20].

In previous investigations, we observed that the MAX phase decomposition in the $Ti_2SnC$ system significantly promotes whisker growth [21]. The extent of phase decomposition induced by chemical-mechanical processing correlates positively with the quantity of whisker growth. This chemical-mechanical phase decomposition, coupled with the electrostatic effects [22-24], are inherently intertwined with the structure and mechanical properties of MAX phases. Therefore, enhancing the mechanical properties of MAX phases and homogenising their structure may provide a more fundamental understanding of whisker growth phenomena within MAX phases.

This study aims to modulate the structure and properties (chemical and electrostatic) of MAX phases by introducing a second element into the M-site of $Ti_2SnC$, thereby suppressing MAX phase decomposition and subsequently inhibiting whisker growth. Taking $Ti_2SnC$ as an example, we introduced varying amounts of vanadium (V) into the M-site. Experimental results indicate that with increasing solid solubility of V in the M-site, the number of whiskers grown on the surface of the samples after identical treatments decreases. When 50% of the M-site atoms are occupied by V atoms, whisker growth is completely suppressed. The combination of experimental and theoretical calculations demonstrates that V solid solution enhances the mechanical properties of the material, strengthens overall chemical bonds, and simultaneously reduces the extent of MAX phase decomposition after identical treatments.

## 2 Method

### Preparation of $(Ti_{1-x}V_x)_2SnC$ (x= 0, 0.25, 0.5, 0.75 and 1)

Commercial Ti powder, V powder (99.99% purity, ⩾300 mesh, Aladdin), Sn powder (99.99% purity, ⩾300 mesh, Aladdin), and graphite powder (99.99% purity, ⩾300 mesh, Aladdin) were employed as raw materials for the synthesis of $(Ti_{1-x}V_x)_2SnC$. Ti powder, V powder, Sn powder, and graphite powder were mixed in a molar ratio of 2(1-$x$):2$x$:1:1 to prepare $(Ti_{1-x}V_x)_2SnC$ ($x$= 0, 0.25, 0.5, 0.75 and 1). The weighed powder was mixed thoroughly (24 h) in a mixer (TURBULA, T2F) and then sintered at fixed temperature for 2 hours in an argon flow to synthesize $(Ti_{1-x}V_x)_2SnC$. The sintering temperatures for five $(Ti_{1-x}V_x)_2SnC$ are displayed in Table 1. The as-synthesized $(Ti_{1-x}V_x)_2SnC$ sample was chemically etched to remove the unreacted elemental Sn. In detail, 5g $(Ti_{1-x}V_x)_2SnC$ were immersed in 100 mL diluted hydrochloric acid (1 mol/L) under magnetic stirring for 4 hours. Then, $(Ti_{1-x}V_x)_2SnC$ was collected by centrifugation and dried in air at 60 °C.

Table 1 The stoichiometric ratio and sintering temperature for $(Ti_{1-x}V_x)_2SnC$.

| Sample | Molar ratio | Temperature (°C) | Product |
| --- | --- | --- | --- |
| S1 | 2Ti/1.1Sn/C | 1330 | $Ti_2SnC$ |
| S2 | 1.5Ti/0.5V/1.1Sn/C | 1300 | $(Ti_{0.75}V_{0.25})_2SnC$ |
| S3 | 1Ti/1V/1.1Sn/C | 1300 | $(Ti_{0.5}V_{0.5})_2SnC$ |
| S4 | 0.5Ti/1.5V/1.1Sn/0.9C | 1100 | $(Ti_{0.25}V_{0.75})_2SnC$ |
| S5 | 2V/1.1Sn/0.9C | 1000 | $V_2SnC$ |

### Mechanochemical decomposition in an oxygen-rich atmosphere

5g of $(Ti_{1-x}V_x)_2SnC$ powder was ball-milled under identical parameters. The ball-milling speed is 650 rpm, and the milling time is 8 hours. To prevent overheating, the milling is paused for 10 minutes after every 30 minutes. After milling, the powder is pressed into discs (Φ16) with a pressure of 800 MPa, and whisker growth is conducted at room temperature.

### Characterization

The phase composition of the milled samples was analyzed by X-ray diffraction (XRD, Haoyuan DX-2700BH) with Cu Kα radiation. The morphology and composition of the milled powders and whiskers were characterized through scanning electron microscopy (SEM, FEI Sirion 200) at a voltage of 20 kV and transmission electron microscopy (TEM, Talos F200X) at a voltage of 200 kV. Tin and titanium oxide content resulting from the decomposition of $Ti_2SnC$ was determined by inductively coupled plasma (ICP, PerkinElmer 8300) analysis.

To measure the Vickers hardness of $(Ti_{1-x}V_x)_2SnC$ solid solution, powder samples were sintered into dense blocks using Spark Plasma Sintering (SPS, FCT-HPD5) system at 1100 °C and 40 MPa for 20 minutes. The resulting $(Ti_{1-x}V_x)_2SnC$ dense blocks were then polished using sandpaper to achieve a smooth, mirror-like surface. The microhardness of the specimens was measured using a Vickers hardness tester (FM-700, Future-Tech) with a load of 500 g applied for 10 s. Hardness values represent the average of ten indentations made on each specimen.

### Theoretical calculation

First-principles calculations based on density functional theory (DFT) [25] were performed using the Vienna Ab initio Simulation Package (VASP) code [26] to investigate the structure and properties of $(Ti_{1-x}V_x)_2SnC$. The projection augmented wave (PAW) method [27] and the generalized gradient approximation implemented in the Perdew-Burke-Ernzerhof (GGA-PBE) method [28] were used for the electron–ion interactions and exchange-correlation functions, respectively. The plane wave cut-off energy of 500 eV and 10×10×2 Monkhorst-Pack grids within Brillouin zone were fixed based on the convergence test. All structures (cell volumes, atom positions, and lattice constants) were fully relaxed until the total energy and forces were less than $1.0×10^{-5}$ eV/atom and 0.02 eV/Å, respectively. After structural optimization, the tetrahedron smearing method [29] was applied for the calculation of the density of states (DOSs), and the band structure was obtained along the explicit path.

## 3 Results and discussion

### Phase analysis

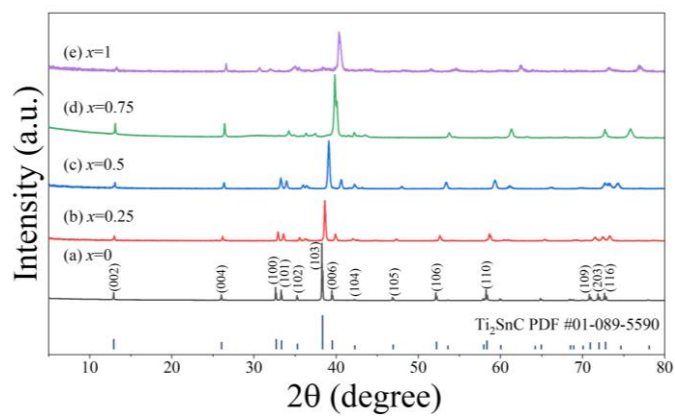

Figure 1 XRD pattern of $(Ti_{1-x}V_x)_2SnC$: (a) $x=0$; (b) $x=0.25$; (c) $x=0.5$; (d) $x=0.75$; (e) $x=1$.

The XRD pattern of the prepared $(Ti_{1-x}V_x)_2SnC$ is shown in Figure 1. The diffraction peaks of all solid solutions exhibit similarities with $Ti_2SnC$, consistent with the characteristic diffraction peaks of the 211-type MAX phase [30]. Due to the smaller atomic radius [31] of the V element (1.35Å) compared to Ti (1.40Å), the lattice of the MAX phase contracts as V atoms gradually dissolve into $Ti_2SnC$ lattice, resulting in a reduction in interplanar spacing. This reduction is manifested as a shift of diffraction peaks towards higher angles in the XRD spectrum. The XRD pattern clearly demonstrates that with the increase of V elements, the strongest peak of the MAX phase (103) shifts towards higher angles [32], transitioning from $Ti_2SnC$ (38.2°) to $V_2SnC$ (40.4°). The prepared solid solution exhibits minimal impurities, except for a small amount of metallic tin present in $V_2SnC$ [33]. Additionally, all synthesized $(Ti_{1-x}V_x)_2SnC$ are composed of clusters formed by the aggregation of nanoparticles ranging in size of several hundred nanometers. Moreover, magnified images show that all solid solutions display a typical layered morphology [34] as shown in Figure 2.

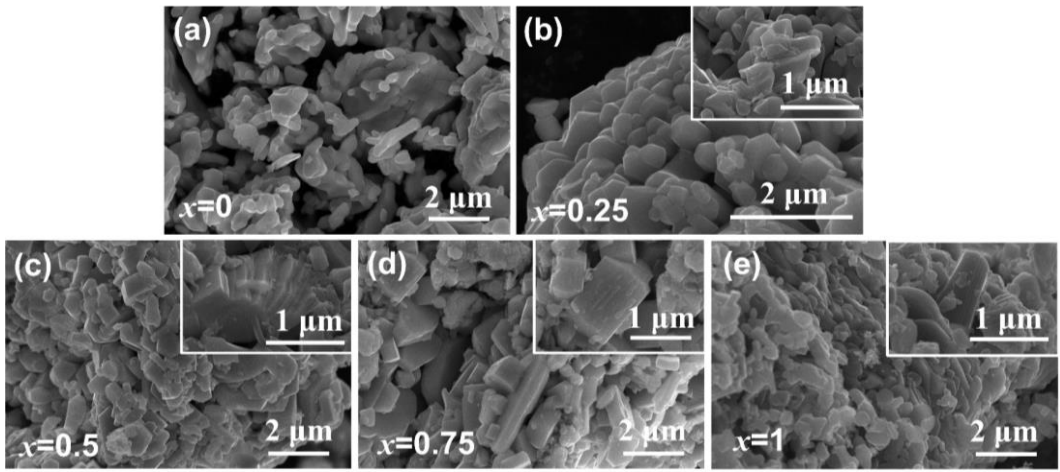

Figure 2 SEM images and magnified morphology of $(Ti_{1-x}V_x)_2SnC$ solid solution: (a) $x=0$; (b) $x=0.25$; (c) $x=0.5$; (d) $x=0.75$; (e) $x=1$.

The actual elemental compositions of all synthesized $(Ti_{1-x}V_x)_2SnC$ were probed by EDS, as shown in Table 2. The actual value of $x$ is consistent with the designed composition of $(Ti_{1-x}V_x)_2SnC$, indicating successful preparation of M-site solid solutions with different content of V elements.

Table 2 Compositional analysis of as-synthesized $(Ti_{1-x}V_x)_2SnC$

| Designed value | Actual composition (at%) | | | | Actual $x$ value |
|---|---|---|---|---|---|
| | Ti | V | Sn | C | |
| $x=0$ | 30.38 | / | 16.02 | 53.60 | $x=0$ |
| $x=0.25$ | 24.33 | 8.35 | 16.21 | 51.11 | $x=0.26$ |
| $x=0.5$ | 15.42 | 14.47 | 15.31 | 54.80 | $x=0.48$ |
| $x=0.75$ | 8.70 | 24.76 | 17.56 | 48.98 | $x=0.74$ |
| $x=1$ | / | 27.91 | 14.21 | 57.88 | $x=1$ |

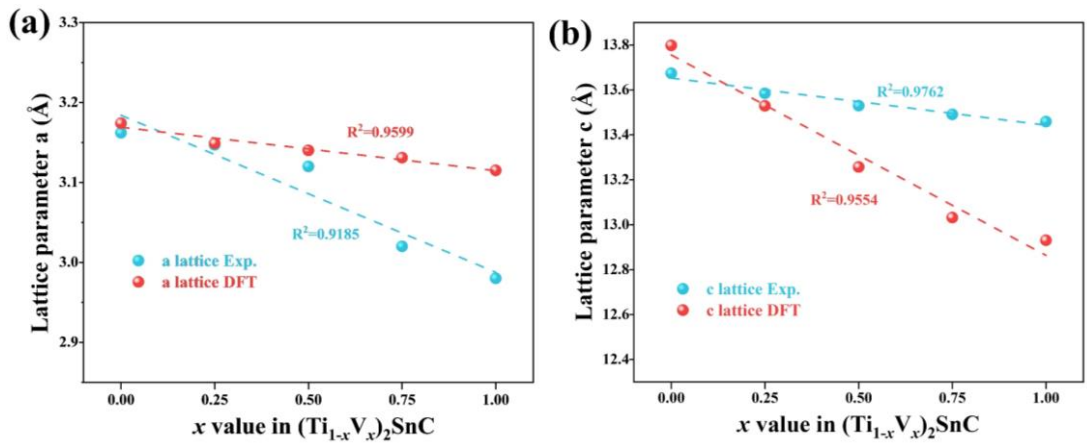

Figure 3 Experimental and theoretical calculations of lattice parameters: (a) lattice parameter a; (b) lattice parameter c.

From the XRD patterns, the crystal structure of the $(Ti_{1-x}V_x)_2SnC$ solid solution remains consistent with its end members, but the lattice parameters underwent changes due to the solid solution of the Vanadium element at the M-site. As depicted in Figure 1, with increasing Vanadium solubility $x$, both the (103) (002) and (100) lattice plane diffraction peaks shift towards higher angles, indicating a monotonic decrease in the lattice parameters a and c with increasing $x$. The experimental and structurally optimized lattice parameters a and c are presented in Figure 3, consistent with the XRD results, where the lattice parameters a and c monotonically decrease with increasing $x$, indicating compliance with Vegard's law [35].

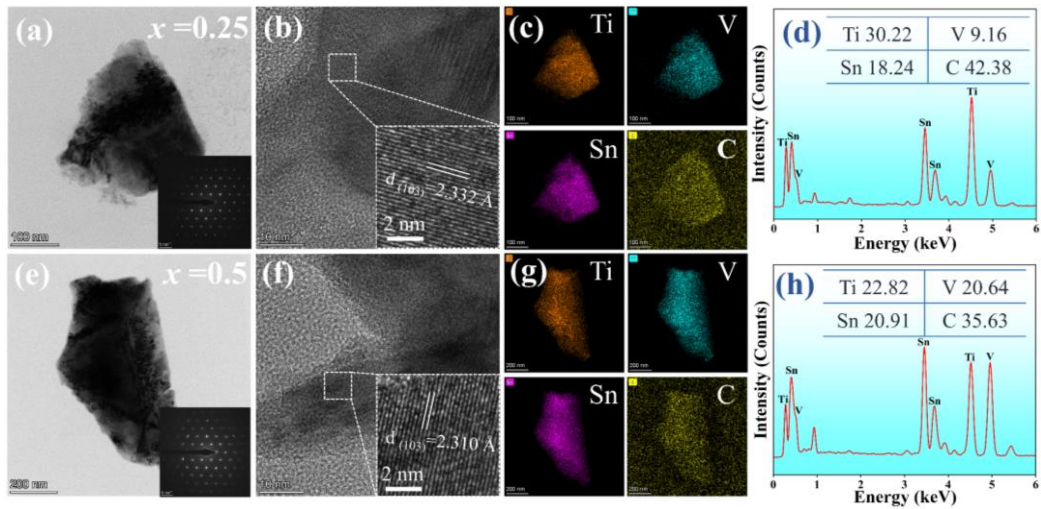

Figure 4 TEM images of $(Ti_{1-x}V_x)_2SnC$ solid solution. (a) TEM image and corresponding SAED pattern (inset) of $(Ti_{0.75}V_{0.25})_2SnC$; (b) High-resolution TEM (HRTEM) image and (103) lattice spacing of $(Ti_{0.75}V_{0.25})_2SnC$; (c) and (d) Elemental distribution maps and elemental composition ratios of $(Ti_{0.75}V_{0.25})_2SnC$; (e) TEM image and SAED pattern (inset) of $(Ti_{0.5}V_{0.5})_2SnC$; (f) HRTEM image and (103) lattice spacing of $(Ti_{0.5}V_{0.5})_2SnC$; (g) and (h) Elemental distribution maps and elemental composition ratios of $(Ti_{0.5}V_{0.5})_2SnC$.

To gain deeper insights into the microstructure and crystal structure of M-site solid solutions, $(Ti_{0.75}V_{0.25})_2SnC$ and $(Ti_{0.5}V_{0.5})_2SnC$ were selected as examples for TEM characterization, as depicted in Figure 4. Figure 4(a) and Figure 4(e) depict the TEM images and SAED patterns (inset) of the prepared $(Ti_{0.75}V_{0.25})_2SnC$ and $(Ti_{0.5}V_{0.5})_2SnC$ respectively. The particle sizes of the $(Ti_{0.75}V_{0.25})_2SnC$ and $(Ti_{0.5}V_{0.5})_2SnC$ solid solutions are in the range of 200-500 nm. The SAED patterns reveal that prepared $(Ti_{1-x}V_x)_2SnC$ belongs to the hexagonal crystal system [36], consistent with the 211 MAX phase. Figure 4(b) and Figure 4(f) depict the HRTEM images of $(Ti_{0.75}V_{0.25})_2SnC$ and $(Ti_{0.5}V_{0.5})_2SnC$ respectively, with interplanar spacings measured by lattice fringe shown in the magnified insets. The interplanar spacing of the (103) plane in $(Ti_{0.75}V_{0.25})_2SnC$ is measured as 2.332 Å, while in $(Ti_{0.5}V_{0.5})_2SnC$, with increasing V content, the interplanar spacing of the (103) plane is 2.310 Å, smaller than that of $Ti_2SnC$ (PDF#089–5590) (2.348 Å) indicating lattice contraction, in close agreement with the (103) interplanar spacing obtained from XRD. EDS mapping reveals the uniform distribution of all constituent elements Ti, V, Sn and C within the $(Ti_{1-x}V_x)_2SnC$ solid solution, and the proportions of each element obtained conform to the predetermined elemental composition ratio.

## Whisker growth

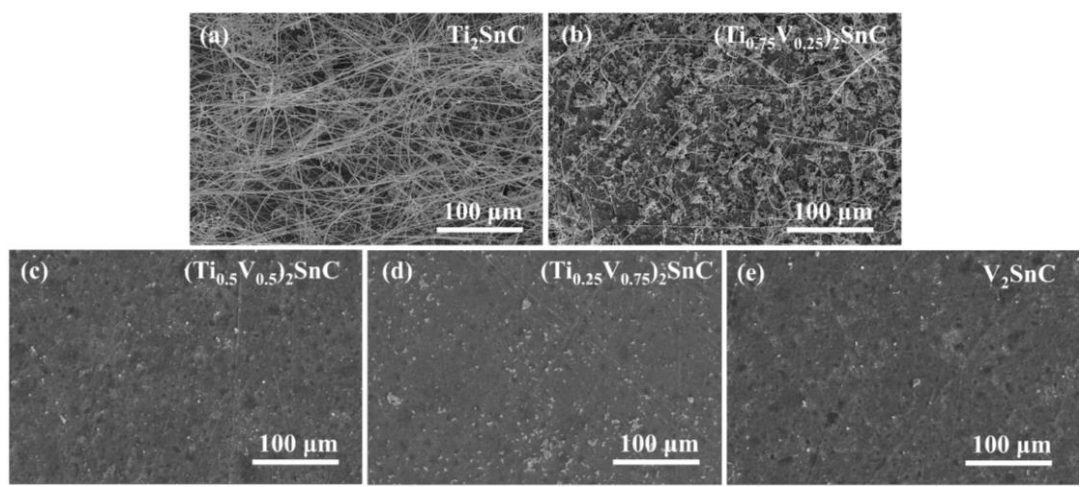

Figure 5 SEM images of whisker growth in $(Ti_{1-x}V_x)_2SnC$ solid solution after treatment with identical parameters: (a) $x=0$; (b) $x=0.25$; (c) $x=0.5$; (d) $x=0.75$; (e) $x=1$.

The samples milled under the same parameters were incubated at room temperature for 7 days. The whisker growth of $(Ti_{1-x}V_x)_2SnC$ is depicted in Figure 5. It can be observed that numerous whiskers grew in $Ti_2SnC$, consistent with our previous studies [37]. After the solid solution of V atoms, the number of whiskers noticeably decreased in the $(Ti_{0.75}V_{0.25})_2SnC$. As the V content continued to increase, the number of whiskers further decreased, only a few whiskers were observed on the sample surface. When the solubility reached $x=0.75$, no whisker growth was observed on the surfaces of the $(Ti_{0.25}V_{0.75})_2SnC$ and $V_2SnC$ samples. It can be demonstrated that the quantity of whisker growth is negatively correlated with the V content. As the V content increases, the number of whiskers gradually decreases in the solid solution. When the solubility reaches $x=0.5$, almost no whisker growth is observed, indicating that the M-site solid solution of V atoms in $Ti_2SnC$ can significantly suppress whisker growth.

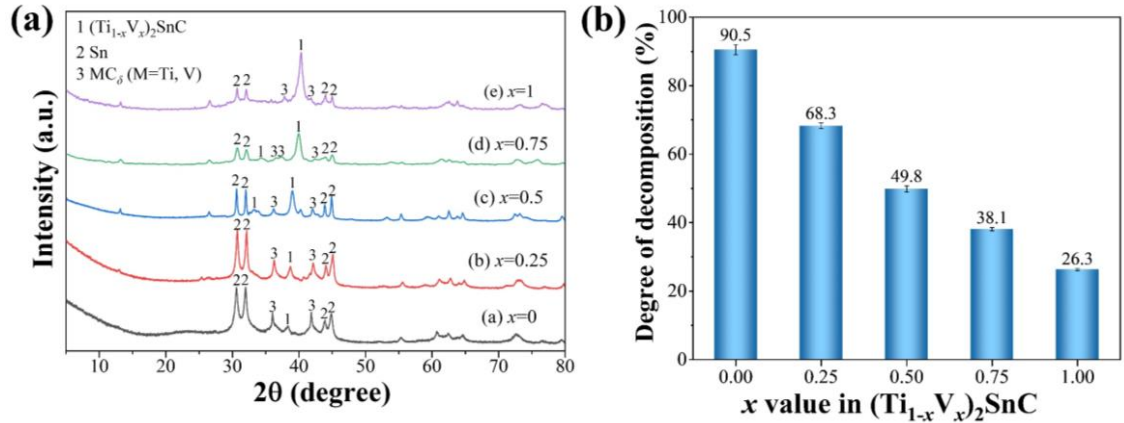

Figure 6 (a) XRD patterns of $(Ti_{1-x}V_x)_2SnC$ solid solution after ball milling with identical parameters; (b) The tin content released from the decomposition of $(Ti_{1-x}V_x)_2SnC$ solid solution measured by Inductively Coupled Plasma (ICP).

To elucidate the mechanism through which V element solid solution impedes the growth of whiskers, we conducted phase characterization of the milled $(Ti_{1-x}V_x)_2SnC$. The XRD pattern (Figure 6a) illustrates that the decomposition products of all $(Ti_{1-x}V_x)_2SnC$ encompass both the undecomposed MAX phases themselves as well as tin metal and metal carbide $MC_δ$ [38]. The black curve represents the XRD pattern of the decomposition products of $Ti_2SnC$, where the intensity of the (103) peak of the MAX phase is notably weakened, nearly disappearing, while the intensities of tin metal and metal carbide are relatively stronger. This signifies a relatively high degree of decomposition of the MAX phase in $Ti_2SnC$. With an increase in V solid solution content, the relative intensity of the strongest peak (103) of the MAX phase rises compared to that of the metal carbide $MC_δ$ and tin metal, indicating a reduction in the extent of decomposition of the solid solution compared to $Ti_2SnC$. Particularly, when the M site is completely replaced by V elements, the predominant decomposition product remains the MAX phase, with only a minor amount of tin and metal carbide peaks generated, suggesting a lower degree of decomposition [21]. This suggests that V solid solution can augment the resistance of MAX phases to mechanochemical decomposition, thereby achieving the inhibition of whisker growth. This discovery corroborates previous reports indicating that whisker growth in MAX phases is directly proportional to the degree of chemical decomposition. The higher the degree of chemical decomposition of MAX phases, the greater the number of whiskers that grow.

To quantitatively characterize the decomposition degree of $(Ti_{1-x}V_x)_2SnC$ solid solution after ball mill, the milled $(Ti_{1-x}V_x)_2SnC$ was immersed in a sufficient amount of dilute hydrochloric acid to dissolve the tin metal generated during decomposition. Subsequently, the supernatant was subjected to ICP testing. The decomposition products consist of metal carbides and tin metal, with the quantity of whisker growth directly correlated to the tin metal content resulting from decomposition. Therefore, we define the percentage of decomposed tin content relative to the tin content in the original MAX phase as the degree of decomposition. As shown in Figure 6(b), it is observed that the decomposition degree of $Ti_2SnC$ reaches 90%, nearly complete decomposition, whereas with increasing V solid solution content, the degree of decomposition gradually decreases, with V$_2$SnC exhibiting a decomposition degree of only 26%. Thus, V element solid solution enhances the resistance of MAX phases to mechanochemical decomposition.

## Mechanism of whisker growth inhibition

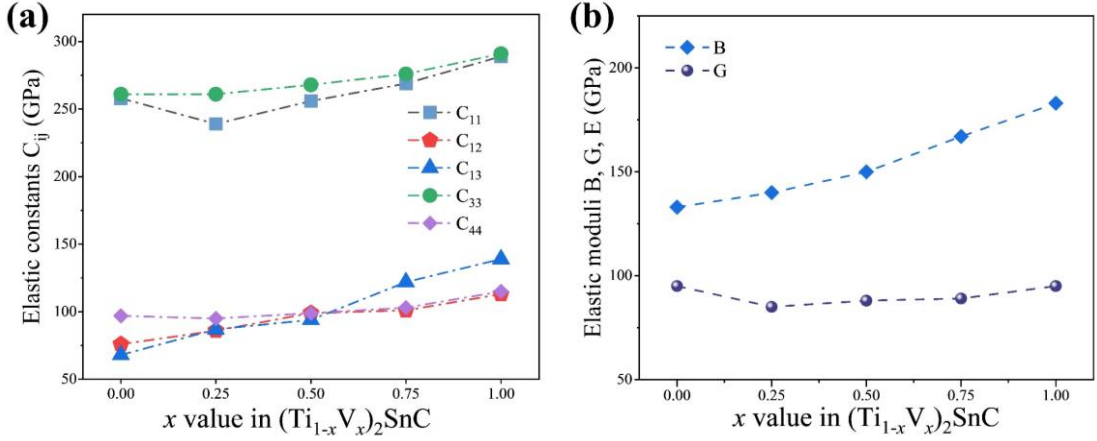

Figure 7 (a) Elastic constants ($C_{11}$, $C_{12}$, $C_{13}$, $C_{33}$, and $C_{44}$) and (b) elastic modulus of (Ti$_{1-x}$V$_x$)$_2$SnC solid solution.

To investigate the impact of V solid solution on the mechanical properties of MAX phases, we computed the elastic properties of (Ti$_{1-x}$V$_x$)$_2$SnC solid solutions, as shown in Figure 7(a). Elastic constants quantify the stiffness of a crystal in response to externally applied strain [39]. The maximum number of independent tensor elements for elastic constants is limited to 21. For different crystal systems, the independent elastic constants are determined by symmetry. As the symmetry of the crystal system increases, the number of independent tensor elements decreases. In hexagonal crystal systems, owing to their high symmetry, there exist only five independent elastic constants, $C_{11}$, $C_{12}$, $C_{13}$, $C_{33}$, and $C_{44}$ [40]. The elastic constant $C_{11}$ measures the elastic stiffness of the crystal under uniaxial stress along the c-axis. $C_{33}$ describes the elastic stiffness of the crystal along the a-axis or b-axis direction. According to the mechanical stability requirements of hexagonal crystals, the elastic constants of all (Ti$_{1-x}$V$_x$)$_2$SnC solid solutions follow the Born elastic stability conditions [41]:

$$C_{44} > 0, C_{11} > |C_{12}|, (C_{11} + 2C_{12})C_{33} > 2C_{13}^2 \tag{1}$$

All principal elastic constants, $C_{11}$ and $C_{33}$, are bigger than the other elastic constants, with $C_{33}$ slightly exceeding $C_{11}$, indicating that the solid solution is less compressible along the c-axis [42]. The proximity of $C_{11}$ and $C_{33}$ values in all solid solutions suggests weak anisotropy. The elastic constants $C_{12}$ and $C_{13}$ represent the stiffness of the crystal along specific directions when subjected to shear stress. Specifically, $C_{12}$ measures the elastic stiffness of the crystal when subjected to shear stress along the a-axis, resulting in strain along the b-axis. $C_{13}$ measures the elastic stiffness of the crystal when subjected to shear stress along the a-axis, resulting in strain along the c-axis [43]. With an increase in the amount of vanadium solid solution, nearly all elastic constants increase, signifying a gradual enhancement in the material's resistance to deformation.

Based on the calculated elastic constants, we further computed the bulk modulus B and shear modulus G of the solid solution using the Voigt-Reuss-Hill approximation [44]. All elastic moduli B and G are displayed in the Figure 7(b). The bulk modulus of the $(Ti_{1-x}V_x)_2SnC$ gradually increases, indicating an enhanced ability to resist volume changes. Furthermore, the bulk modulus reflects the overall chemical bond strength of the material. With the solid solution of vanadium solute, the overall chemical bonding within the MAX phase strengthens [45]. The shear modulus G refers to the material's resistance to shear deformation, indicating its ability to resist plastic deformation. The shear modulus of solid solutions is lower than that of $Ti_2SnC$, suggesting increased plasticity of the material. The combination of higher bulk modulus and lower shear modulus is indicative of the greater damage tolerance exhibited by MAX phase materials [3].

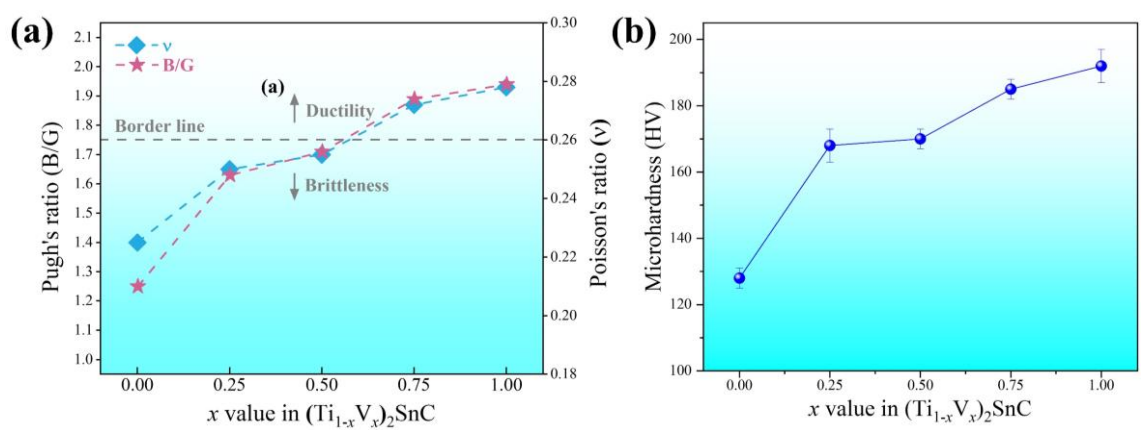

Figure 8 (a) Pugh's ratio (B/G), Poisson's ratio (ν), and (b) experimentally determined Vickers hardness of $(Ti_{1-x}V_x)_2SnC$ solid solution.

To comprehensively assess the mechanical properties of $(Ti_{1-x}V_x)_2SnC$ solid solution, the B/G ratio, Poisson's ratio, and experimentally measured Vickers hardness of the MAX phase were calculated and measured, as shown in Figure 8. Pugh's ratio is defined as the ratio of bulk modulus B to shear modulus G. When the B/G ratio is greater than 1.75, the material is a ductile phase, and when the B/G ratio is less than 1.75, the material appears as a brittle phase [46]. It can be observed that the B/G ratio of $Ti_2SnC$ is 1.4, indicative of a typical brittle material. As the content of vanadium element at the M-site increases, the B/G ratio of $(Ti_{1-x}V_x)_2SnC$ solid solution increases. Specifically, the B/G ratio of $(Ti_{0.75}V_{0.25})_2SnC$ is 1.65, and that of $(Ti_{0.5}V_{0.5})_2SnC$ is 1.70, indicating an increase in ductility. With further increase in the solubilized vanadium content, the B/G ratio of $(Ti_{0.25}V_{0.75})_2SnC$ reaches 1.87, and that of $V_2SnC$ is 1.93, both exceeding 1.75, demonstrating ductile behavior. Additionally, Poisson's ratio can also gauge the ductility of materials, and it is defined as:

$$\nu = \frac{3B - 2G}{6B + 2G} \quad (2)$$

When the Poisson's ratio exceeds 0.26, the material is considered ductile, whereas when it is below 0.26, the material is brittle [5]. As shown in Figure 8(a), it can be observed that the Poisson's ratio and B/G ratio of $(Ti_{1-x}V_x)_2SnC$ solid solution exhibit a similar trend. That is, with the increase in the solubilized quantity of vanadium (V) element at the M-site, the Poisson's ratio of $(Ti_{1-x}V_x)_2SnC$ solid solution gradually increases, indicating a transition from brittleness to ductility in the material.

The prepared $(Ti_{1-x}V_x)_2SnC$ solid solution powders were subjected to Spark Plasma Sintering (SPS) to sinter the powders into dense bulk forms, and Vickers hardness testing was conducted on the $(Ti_{1-x}V_x)_2SnC$ bulk samples. Figure 8(b) depicts the Vickers hardness measurements of $(Ti_{1-x}V_x)_2SnC$, where the Vickers hardness of $Ti_2SnC$ is 128 HV. The hardness of materials with vanadium (V) element solubilization and interstitial $V_2SnC$ are both enhanced, attributed to the solid solution strengthening at the M-site. The chemical bonds within the MAX phase are strengthened, leading to improved mechanical properties. Some researchers have employed Mulliken population theory to calculate the Vickers hardness of $M_2SnC$, and theoretical calculations similarly indicate that the Vickers hardness of $V_2SnC$ is higher than that of $Ti_2SnC$ [47].

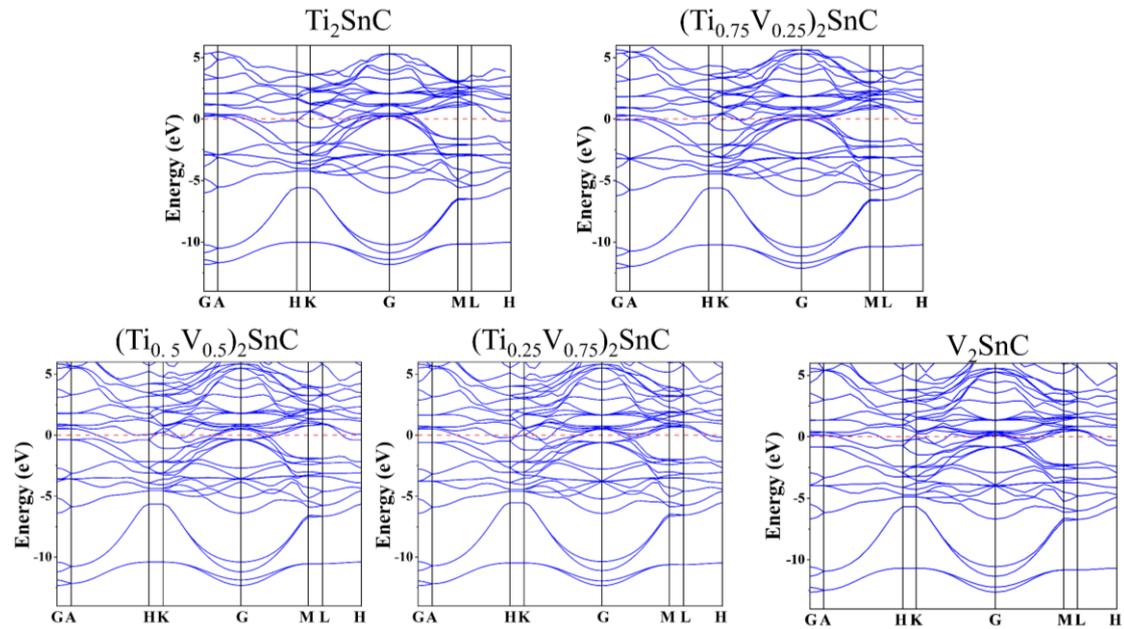

Figure 9 Electronic band and structure of $(Ti_{1-x}V_x)_2SnC$: (a) $x=0$; (b) $x=0.25$; (c) $x=0.5$; (d) $x=0.75$; (e) $x=1$.

Electronic structures play a crucial role in understanding material properties at the microscopic level [43]. Based on structural optimization, the band structures of $(Ti_{1-x}V_x)_2SnC$ solid solutions were calculated at various points along high-symmetry directions in the Brillouin zone, as shown in Figure 9. At the Fermi level, there is a noticeable overlap between the valence and conduction bands for all five MAX phases, indicating the absence of a significant band gap. This suggests that the five $(Ti_{1-x}V_x)_2SnC$ solid solutions exhibit metallic properties.

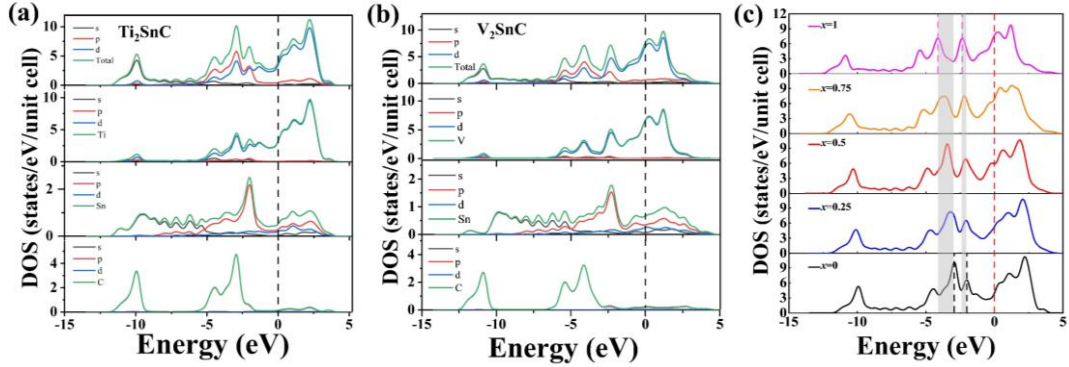

Figure 10 Density of state of (a) Ti$_2$SnC and (b) V$_2$SnC; (c) total density of state for different V content $x$.

The electronic Density of States (DOS) can reflect the bonding situations between atoms, revealing the chemical bonding properties and structural characteristics of compounds [32]. We computed the Total Density of States (TDOS) and Partial Density of States (PDOS) for both Ti$_2$SnC and V$_2$SnC MAX phases, as depicted in Figure 10. The calculated values of the DOS at the Fermi level for Ti$_2$SnC and V$_2$SnC are both greater than zero. In Ti$_2$SnC, the DOS at the Fermi level is mainly contributed by Ti 3d orbitals, while in V$_2$SnC, it is primarily contributed by V 3d orbitals. For the (Ti$_{1-x}$V$_x$)$_2$SnC solid solution, states near and above the Fermi level are attributed to intermetallic d-d interactions and antibonding states. The lower valence band (LVB) is contributed by the C 2s orbital, while the middle valence band consists of Ti 3d/V 3d and C 2p orbitals, corresponding to the energy overlap peaks of Ti 3d/V 3d and C 2p. The upper valence band (UVB) is composed of Ti 3d/V 3d and Sn 5p orbitals, corresponding to the energy overlap peaks of Ti 3d/V 3d and Sn 5p.

In Ti$_2$SnC, there is an energy overlap peak around -3 eV for Ti-3d and C 2p orbitals, indicating a strong s-p hybridization between Ti and C atoms. Meanwhile, the hybridization peak energy between Ti 3d and Sn 5p orbitals is around -2 eV. The energy of the Ti 3d and C 2p hybridization peak is lower than that of the Ti 3d and Sn 5p hybridization peak (-2 eV), suggesting that the Ti-C bond is stronger than the Ti-Sn bond [48], consistent with the conclusion that in MAX phases, M-X bonds are stronger than M-A bonds (M= Ti/V). Similar conclusions can be drawn for V$_2$SnC, where the energy of the hybridization peak between V 3d and C 2p orbitals (-4 eV) is lower than that of the hybridization peak between V 3d and Sn 5p orbitals (-2.3 eV), indicating again that the V-C bond is stronger than the V-Sn bond.

The energy of the hybridization peak between Ti 3d and C 2p orbitals (-3 eV) is lower than that between V 3d and C 2p orbitals (-4 eV). This suggests that the Ti-C bond is stronger than the V-C bond. Similarly, the energy of the hybridization peak between Ti 3d and Sn 5p orbitals (-2 eV) is lower than that between V 3d and Sn 5p orbitals (-2.3 eV), indicating that the Ti-Sn bond is stronger than the V-Sn bond. Overall, the chemical bond strength in V$_2$SnC is stronger than that in Ti$_2$SnC, which is consistent with the results obtained from the bulk modulus measurements. We stack the TDOS for five solid solutions together to facilitate the observation of energy shift trends, as shown in Figure 10(c). As the V solid solubility $x$ increases, the overlap peak energies of the

M 3d/C 2p bond and M 3d/Sn 5p bond gradually shift to lower energy levels. This indicates that with the incorporation of the V element, both the M-C and M-Sn bonds are strengthened, and these enhancements are positively correlated with the V solid solubility. In V$_2$SnC, it exhibits the lowest overlap energies between Ti 3d/V 3d and Sn 5p, as well as Ti 3d/V 3d and C 2p; in terms of bulk modulus, V$_2$SnC has the highest bulk modulus, confirming these observations.

Table 3 The bond lengths of (Ti$_{1-x}$V$_x$)$_2$SnC solid solution after structural optimization.

| x value | Ti-C (Å) | V-C (Å) | Ti-Sn (Å) | V-Sn (Å) |
|---|---|---|---|---|
| x=0 | 2.143 | / | 2.968 | / |
| x=0.25 | 2.142 | 2.074 | 2.961 | 2.937 |
| x=0.5 | 2.139 | 2.072 | 2.952 | 2.924 |
| x=0.75 | 2.139 | 2.066 | 2.946 | 2.918 |
| x=1 | / | 2.056 | / | 2.912 |

Based on the structural optimization of (Ti$_{1-x}$V$_x$)$_2$SnC solid solution, we measured the M-C and M-Sn bond lengths in the optimized models. As shown in Table 3, in Ti$_2$SnC, the Ti-C bond length is 2.146 Å, and the Ti-Sn bond length is 2.968 Å. With an increase of vanadium content, the Ti-C bond length gradually decreases, partially replaced by shorter V-C bonds, and similarly, the Ti-Sn bond length shortens, partially replaced by shorter V-Sn bonds. This trend continues until the M-site is fully substituted by vanadium, at which point the V-C bond length measures 2.056 Å, and the V-Sn bond length measures 2.912 Å. The shorter the bond length, the stronger the bond energy [34]. Both M-C and M-Sn bonds are strengthened concurrently, leading to superior mechanical properties of the MAX phase after vanadium element solubilization.

Another scenario worth considering is based on the electrostatic theory of whisker growth [23]. According to this theory, significant electric fields can induce the formation of metal whiskers by making them energetically favorable. These electric fields may be externally applied or intrinsic to the materials due to uncompensated electric charges at interfaces, grain boundaries, local variations in chemical compositions, and other factors [22-24].

The essence of electrostatic theory is illustrated in Figure 11 sketching random charge patches on a metal surface. Depending on a particular combination of patches, their induced electric field strength $E$ distance $h$ from the surface will be different. A metal filament of length $h$ will be correspondingly polarized and the polarization energy $-pE$, where $p$ is the dipole moment, constitutes energy gain illustrated in Figure 12 thus promoting whiskers development at favorable locations.

Referring to the published quantitative analyses [22-24, 49], there are two general conclusions of relevance to this project: 1) Surface tension is a parameter discriminating between the differences in propensities for whisker growth between different alloys, and 2) The surface charge density (or, which is equivalent, its electric field strength) is a physical quantity affecting whisker developments.

The surface tension anticorrelates with the material hardness or microhardness [50], which characteristics thus become indicative of whisker propensity. As for the surface charge density, the concept of charge patches on metal surfaces was put forward [51], almost as long time ago as the first whisker observations.

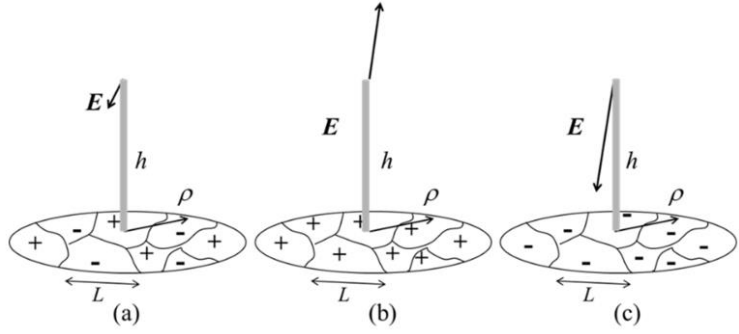

Figure 11 Random charge patch configurations on a metal surface inducing random electric fields.

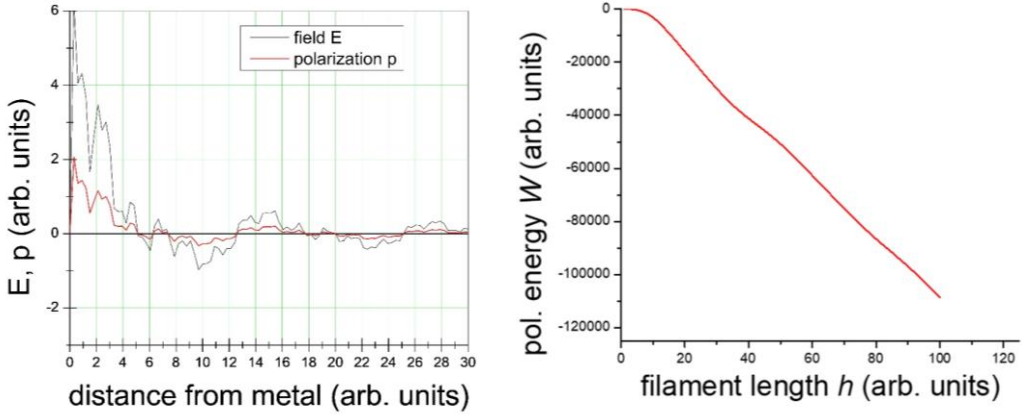

Figure 12 Left: Electric field and polarization vs distance from metal surface in the random patch model. Right: The corresponding polarization energy constituting whisker promoting gain.

It was driven by data on electron emission. There has been a considerable effort in developing the charge patch concept [23], not related to metal propensity for whisker growth. According to that concept, the surface remains neutral, so that its negative and positive patch charges mutually balance each other overall yet can create significant local effects. The characteristic patch size in Figure 12 is assumed to be in the micron or sub-micron ranges.

We should note as well that it is not necessarily directly related to electric charges deposited on a material face: they can reflect a variety of conditions originating in its depth. Several such factors should be noted: impurity related chemical composition fluctuations; different grain orientations rendering surfaces with work functions different by several tenths of electron-volt, internal stresses and stress gradients producing electrical nonuniformities through the deformation potential mechanism, intermetallic compounds, and recrystallization [23].

As a useful analytical guide, consider the following expression for the whisker nucleation barrier [24],

$$W = \sigma R \sqrt{\frac{\sigma \Lambda R}{\epsilon E^2}} \quad \text{with} \quad \Lambda = \ln\left(\frac{2h}{R}\right) - 1 \qquad (3)$$

Here $\sigma$ is the surface tension, $\epsilon$ is the dielectric permittivity, $E$ is the local field strength, $h$ is the whisker length and $R$ is the whisker radius. The characteristic whisker nucleation time is then estimated as $\tau = \tau_0 \exp(-W/kT)$ where $\tau_0 \sim 10^{-13}$ s, and $kT$ has its standard meaning. The latter expression shows how whisker probability is exponentially sensitive the field strength and surface tension.

The electrostatic theory of metal whiskers provides a unifying framework for understanding our observations. Indeed,
1) Increasing the V content in MAX alloys suppresses the degree of decomposition, hence, decreasing the electric charge density on the composite interfaces, which leads to weaker electric fields governing the whisker nucleation barrier $W$ and exponential suppression of whisker probability.
2) According to the above discussion, our observed microhardness will increase with V content pointing towards larger surface tensions and thus exponential decrease in whisker probability/concentration.

## 4 Conclusion

In this study, we introduced varying amounts of vanadium elements into the M-site of $Ti_2SnC$. As the solubilized quantity of V element increased, the number of whiskers decreased, and the degree of chemical decomposition reduced gradually. Computational studies using Density Functional Theory were conducted to investigate the structure and mechanical properties of $(Ti_{1-x}V_x)_2SnC$. Experimental and DFT-derived lattice constants a and c exhibited a linear decrease with increasing V solubilization content $x$, in accordance with Vegard's law. Experimental and DFT results demonstrated that V element solubilization significantly enhanced the Vickers hardness and toughness of the MAX phase. The DOS indicated stronger V-C bonds compared to Ti-C bonds and stronger V-Sn bonds compared to Ti-Sn bonds. Additionally, structural optimization revealed that V element solubilization reinforced the overall chemical bonds of the material. The enhancement in mechanical properties, combined with electrostatic effects, strengthened the resistance of $(Ti_{1-x}V_x)_2SnC$ against chemical decomposition, thereby suppressing whisker growth.

## CRediT authorship contribution statement

**Haifeng Tang:** Writing – original draft, Writing – review & editing Data curation, Methodology, Investigation. **Xiaodan Yin:** Validation, Formal analysis, Investigation. **Peigen Zhang:** Validation, Formal analysis, Funding acquisition, Resources, Supervision, and Writing, reviewing, and editing. **Victor Karpov:** Methodology, Formal analysis, Validation. **Vamsi Borra:** Methodology, Formal analysis, Validation. **Zhihua Tian:** Methodology, Formal analysis, Validation. **Jianxiang Ding:** Formal analysis; Validation, Supervision. **ZhengMing Sun:** Validation, Funding acquisition, Supervision, and Writing, reviewing, and editing.

## Declaration of competing interest

The authors declare that they have no known competing financial interests or personal relationships that could have appeared to influence the work reported in this paper.

## Data availability

Data will be made available on request.

## Acknowledgement

This work was financially supported by the National Natural Science Foundation of China (52171033).

# 5 Reference

[1] M. Dahlqvist, M.W. Barsoum, J. Rosen, MAX phases – Past, present, and future, Mater. Today 72 (2024) 1-24.

[2] M. Sokol, V. Natu, S. Kota, M.W. Barsoum, On the Chemical Diversity of the MAX Phases, Trends Chem. 1(2) (2019) 210-223.

[3] A.C. Yang, Y.H. Duan, L.K. Bao, M.J. Peng, L. Shen, Elastic properties, tensile strength, damage tolerance, electronic and thermal properties of $TM_3AlC_2$ (TM = Ti, Zr and Hf) MAX phases: A first-principles study, Vacuum 206 (2022) 111497.

[4] Y.H. Niu, S. Fu, K.B. Zhang, B. Dai, H.B. Zhang, S. Grasso, C.F. Hu, Synthesis, microstructure, and properties of high purity $Mo_2TiAlC_2$ ceramics fabricated by spark plasma sintering, J. Am. Ceram. Soc. 9(6) (2020) 759-768.

[5] X.J. Duan, Z. Fang, T. Yang, C.Y. Guo, Z.K. Han, D. Sarker, X.M. Hou, E.H. Wang, Maximizing the mechanical performance of $Ti_3AlC_2$-based MAX phases with aid of machine learning, J. Am. Ceram. Soc. 11(8) (2022) 1307-1318.

[6] D.D. Wang, W.B. Tian, C.J. Lu, J.X. Ding, Y.F. Zhu, M. Zhang, P.G. Zhang, Z.M. Sun, Comparison of the interfacial reactions and properties between $Ag/T_3AlC_2$ and $Ag/Ti_3SiC_2$ electrical contact materials, J. Alloys Compd. 857 (2021) 157588.

[7] X.L Wu, C.Z Wu, X.P Wei, W.J Sun, C.J Ma, Y.D Zhang, G.G Li, L.M Chen, D.D Wang, Influence of nano-mechanical evolution of $Ti_3AlC_2$ ceramic on the arc erosion resistance of Ag-based composite electrical contact material., J. Adv. Ceram. 13(2) (2024) 176-188.

[8] J. Gonzalez-Julian, I. Kraleva, M. Belmonte, F.B. Jung, T. Gries, R. Bermejo, Multifunctional performance of $Ti_2AlC$ MAX phase/2D braided alumina fiber laminates, J. Am. Ceram. Soc. 105(1) (2022) 120-130.

[9] Q.Y. Tan, W.M. Zhuang, M. Attia, R. Djugum, M.X. Zhang, Recent progress in additive manufacturing of bulk MAX phase components: A review, J. Mater. Sci. Technol. 131 (2022) 30-47.

[10] M.W. Barsoum, L. Farber, Room-temperature deintercalation and self-extrusion of Ga from $Cr_2GaN$, Science 284(5416) (1999) 937-939.

[11] C.J. Lu, Y.S. Liu, J. Fang, Y. Zhang, P.G. Zhang, Z.M. Sun, Isotope study reveals atomic motion mechanism for the formation of metal whiskers in MAX phase, Acta Mater. 203 (2021) 116475.

[12] S. Li, Y.S. Liu, P.G. Zhang, Y. Zhang, C.J. Lu, L. Pan, J.X. Ding, Z.M. Sun, Interface energy-driven indium whisker growth on ceramic substrates, J. Mater. Sci. Mater. Electron. 32(12) (2021) 16881-16888.

[13] P.G. Zhang, J.X. Ding, Y.S. Liu, L. Yang, W.B. Tian, J. Ouyang, Y.M. Zhang, Z.M. Sun, Mechanism and mitigation of spontaneous Ga whisker growth on $Cr_2GaC$, Sci. China Technol. Sci. 63(3) (2020) 440-445.

[14] Y.A. Ren, Z.H. Tian, Y. Zhang, F.S. Wu, H. Xie, Q.Q. Zhang, P.G. Zhang, Z.M. Sun, In-Situ Growth of ZnO Whiskers on $Ti_2ZnC$ MAX Phases, Materials 16(10) (2023) 3610.

[15] H.L. Cobb, Cadmium whiskers, Mon. Rev. Am. Electroplat. Soc 33(1) (1946) 28-30.

[16] NASA, Tin Whisker (and Other Metal Whisker) Homepage, 2023. https://nepp.nasa.gov/whisker/index.html.
[17] L.J. Qiao, J.Q. Bi, G.D. Liang, C. Liu, Z.Z. Yin, Y. Yang, H.Y. Wang, S.Y. Wang, M.M. Shang, W.L. Wang, Synthesis and electromagnetic wave absorption performances of a novel $(Mo_{0.25}Cr_{0.25}Ti_{0.25}V_{0.25})_3AlC_2$ high-entropy MAX phase, J. Mater. Sci. Technol. 137 (2023) 112-122.
[18] T. Lapauw, D. Tytko, K. Vanmeensel, S.G. Huang, P.P. Choi, D. Raabe, E.N. Caspi, O. Ozeri, M.T. Baben, J.M. Schneider, K. Lambrinou, J. Vleugels, $(Nb_x, Zr_{1-x})_4AlC_3$ MAX Phase Solid Solutions: Processing, Mechanical Properties, and Density Functional Theory Calculations, Inorg. Chem. 55(11) (2016) 5445-5452.
[19] L.S. Qu, G.P. Bei, B. Stelzer, H. Ruess, J.M. Schneider, D.X. Cao, S. van der Zwaag, W.G. Sloof, Synthesis, crystal structure, microstructure and mechanical properties of $(Ti_{1-x}Zr_x)_3SiC_2$ MAX phase solid solutions, Ceram. Int 45(1) (2019) 1400-1408.
[20] X. Wei, L. Li, F. Liu, L. Fan, Y. Wu, Synthesis, microstructure, and mechanical properties of MAX phase $Ti_2GaC$ ceramics with V doping, Ceram. Int 50(9) (2024) 15806-15820.
[21] Q.Q. Zhang, Z.H. Tian, P.G. Zhang, Y. Zhang, Y.S. Liu, W. He, L. Pan, Y. Liu, Z.M. Sun, Rapid and massive growth of tin whisker on mechanochemically decomposed $Ti_2SnC$, Mater. Today Commun. 31 (2022) 103466.
[22] V.G. Karpov, Electrostatic Theory of Metal Whiskers, Phys. Rev. Appl. 1(4) (2014) 044001.
[23] D. Shvydka, V.G. Karpov, Surface parameters determining a metal propensity for whiskers, J. Appl. Phys. 119(8) (2016) 085301.
[24] V. Borra, D.G. Georgiev, V.G. Karpov, D. Shvydka, Microscopic Structure of Metal Whiskers, Phys. Rev. Appl. 9(5) (2018) 054029.
[25] P. Hohenberg, W. Kohn, Inhomogeneous electron gas, Phys. Rev. B 136(3B) (1964) B864.
[26] J. Hafner, Ab-initio simulations of materials using VASP: Density-functional theory and beyond, J. Comput. Chem. 29(13) (2008) 2044-2078.
[27] P.E. Blochl, Projector augmented-wave method, Phys. Rev. B 50(24) (1994) 17953-17979.
[28] J.P. Perdew, K. Burke, M. Ernzerhof, Generalized gradient approximation made simple Phys. Rev. Lett. 78(7) (1997) 1396-1396.
[29] P.E. Blochl, O. Jepsen, O.K. Andersen, Improved tetrahedron method for Brillouin-zone integrations, Phys. Rev. B 49(23) (1994) 16223-16233.
[30] L. Fu, W. Xia, MAX Phases as Nanolaminate Materials: Chemical Composition, Microstructure, Synthesis, Properties, and Applications, Adv. Eng. Mater. 23(4) (2021) 2001191.
[31] J.C. Slater, Atomic Radii in Crystals, J. Chem. Phys. 41(10) (1964) 3199-3204.
[32] Z.H. Tian, B.Z. Yan, F.S. Wu, J.W. Tang, X.Q. Xu, J. Liu, P.G. Zhang, Z.M. Sun, Synthesis of $Ti_2(In_xAl_{1-x})C$ (x=0-1) solid solutions with high-purity and their properties, J. Eur. Ceram. Soc. 43(14) (2023) 5915-5924.
[33] Q. Xu, Y.C. Zhou, H.M. Zhang, A.N. Jiang, Q.Z. Tao, J. Lu, J. Rosén, Y.H. Niu, S.

Grasso, C.F. Hu, Theoretical prediction, synthesis, and crystal structure determination of new MAX phase compound V$_2$SnC, J. Am. Ceram. Soc. 9(4) (2020) 481-492.

[34] Z.H. Tian, P.G. Zhang, W.W. Sun, B.Z. Yan, Z.M. Sun, Vegard's law deviating Ti$_2$(Sn$_x$Al$_{1-x}$)C solid solution with enhanced properties, J. Adv. Ceram 12(8) (2023) 1655-1669.

[35] Y.X. Zhou, Y.C. Rao, L.T. Zhang, S.H. Ju, H. Wang, Machine-learning prediction of Vegard's law factor and volume size factor for binary substitutional metallic solid solutions, Acta Mater. 237 (2022) 118166.

[36] Z.J. Wang, Y.F. Ma, K. Sun, Q. Zhang, C. Zhou, P.Z. Shao, Z.Y. Xiu, G.H. Wu, Enhanced ductility of Ti$_3$AlC$_2$ particles reinforced pure aluminum composites by interface control, Mater. Sci. Eng. A 832 (2022) 142393.

[37] H.F. Tang, B.Z. Yan, P.G. Zhang, X.D. Yin, Z.H. Tian, S. Das Mahapatra, W. Zheng, J.W. Tang, Z.M. Sun, Controlling tin whisker growth via oxygen-mediated decomposition of Ti$_2$SnC, J. Mater. Sci. 59(5) (2024) 1958-1967.

[38] Z.H. Tian, X.Q. Xu, J.W. Tang, Q.Q. Zhang, F.S. Wu, P.G. Zhang, J. Liu, Z.M. Sun, Large-scale preparation of nano-sized carbides and metal whiskers via mechanochemical decomposition of MAX phases, Int. J. Appl. Ceram. Technol. 20(2) (2023) 823-832.

[39] M. de Jong, W. Chen, T. Angsten, A. Jain, R. Notestine, A. Gamst, M. Sluiter, C.K. Ande, S. van der Zwaag, J.J. Plata, C. Toher, S. Curtarolo, G. Ceder, K.A. Persson, M. Asta, Charting the complete elastic properties of inorganic crystalline compounds, Sci. Data 2 (2015) 150009.

[40] L. Fast, J.M. Wills, B. Johansson, O. Eriksson, Elastic constants of hexagonal transition metals: Theory, Phys. Rev. B 51(24) (1995) 17431-17438.

[41] F. Mouhat, F.-X. Coudert, Necessary and sufficient elastic stability conditions in various crystal systems, Phys. Rev. B 90(22) (2014) 224104.

[42] W.B. Gauster, I.J. Fritz, Pressure and temperature dependences of the elastic constants of compression‐annealed pyrolytic graphite, J. Appl. Phys. 45(8) (1974) 3309-3314.

[43] M. Ali, Z. Bibi, M.W. Younis, K. Majeed, M.A. Iqbal, First-principles investigation of structural, mechanical, and optoelectronic properties of Hf$_2$AX (A = Al, Si and X = C, N) MAX phases, J. Am. Ceram. Soc. 107(4) (2024) 2679-2692.

[44] L. Zuo, M. Humbert, C. Esling, Elastic properties of polycrystals in the Voigt-Reuss-Hill approximation, J. Appl. Crystallogr. 25 (1992) 751-755.

[45] Q.Q. Zhang, Y.C. Zhou, X.Y. San, W.B. Li, Y.W. Bao, Q.G. Feng, S. Grasso, C.F. Hu, Zr$_2$SeB and Hf$_2$SeB: Two new MAB phase compounds with the Cr$_2$AlC-type MAX phase (211 phase) crystal structures, J. Am. Ceram. Soc. 11(11) (2022) 1764-1776.

[46] D. Behera, A. Dixit, A. Azzouz-Rached, A. Bentouaf, M.F. Rahman, H. Albalawi, A. Bouhenna, E. Yousef, R. Sharma, Prediction of new MAX phase Zr$_2$MSiC$_2$ (M = Ti, V) compounds as a promising candidate for future engineering: DFT calculations, Mater. Sci. Eng. B 301 (2024) 117141.

[47] M.A. Hadi, M. Dahlqvist, S.R.G. Christopoulos, S.H. Naqib, A. Chroneos, A.K.M.A. Islam, Chemically stable new MAX phase V$_2$SnC: a damage and radiation tolerant TBC material, RSC Adv. 10(71) (2020) 43783-43798.

[48] Z.Q. Li, E.X. Wu, K. Chen, X.D. Wang, G.X. Chen, L.J. Miao, Y.M. Zhang, Y.J. Song, S.Y. Du, Z.F. Chai, Q. Huang, Chalcogenide MAX phases $Zr_2Se(B_{1-x}Se_x)$ (x=0-0.97) and their conduction behaviors, Acta Mater. 237 (2022) 118183.

[49] M. Killefer, V. Borra, A. Al-Bayati, D.G. Georgiev, V.G. Karpov, E.I. Parsai, D. Shvydka, Whisker growth on Sn thin film accelerated under gamma-ray induced electric field, J. Phys. D: Appl. Phys. 50(40) (2017) 405302.

[50] K.A. Osintsev, I.A. Komissarova, S.V. Konovalov, S.V. Voronin, X. Chen, The influence of electrical potential on the mechanical properties of commercially pure titanium, Lett. Mater. 10(4) (2020) 512-516.

[51] V. Borra, S. Itapu, D.G. Georgiev, Sn whisker growth mitigation by using NiO sublayers, J. Phys. D: Appl. Phys. 50(47) (2017) 475309.